


\def\mydate{June 22, 1994}

\hsize=14.cm
\vsize=21.cm   
\hoffset=1.2cm
\voffset=1.2cm



\baselineskip=13pt plus 1pt minus 1pt  
\tolerance=1000
\parskip=0pt
\parindent=15pt

\def\normal{\baselineskip=13pt plus 1pt minus 1pt}

\nopagenumbers

\newdimen\pagewidth  \newdimen\pageheight
\pagewidth=\hsize  \pageheight=\vsize

\font\tenss=cmss10

\font\bfBig=cmb10  scaled\magstep2

\font\fiverm=cmr5

\font\sevenrm=cmr7
\font\eightrm=cmr8
\font\ninerm=cmr9
\font\tenrm=cmr10

\font\nineit=cmti9
\font\eightit=cmti8
\font\sevenmit=cmmi7
\font\sixmit=cmmi6

\font\eightsl=cmsl8

\font\ninebf=cmbx9

\font\tencal=cmsy10
\font\eightcal=cmsy8


\def\funR{\hbox{\tencal\char'74}}

\def\greekPi{{\char'5}}

\def\smallomega{\hbox{\sevenmit\char'41}}
\def\smallsim{\hbox{\eightcal\char'30}}
\def\smallcdot{\hbox{\eightcal\char'1}}
\def\smallminus{\hbox{\eightcal\char'0}}

\def\smallhbar{{\eightit h} \kern-.8em\hbox{\eightrm\char'26}}

\def\myref#1{[{#1}]~}
\def\myreff#1{[{#1}]}


\def\hspace{~\hskip 1cm ~}

\def\big{\displaystyle \strut }

\def\N{\kappa}

\def\vx{ {\vec x} }

\def\vq{ {\vec q}\, }
\def\vp{ {\vec p}\, }

\def\bz{ {\bar z} }

\def\x{ {\bf x} }
\def\y{ {\bf y} }

\def\ZZ{ \hbox{\tenss Z} \kern-.4em \hbox{\tenss Z} }

\def\L{ {\cal L} }

\def\Lag{ {\cal N} }
\def\E{ {\cal E} }
\def\D{ {\cal D} }

\def\ep{\epsilon}
\def\eps{\varepsilon^{\mu\nu\rho}}
\def\d{\partial}

\def\tr{ {\,{\rm tr}\,} }
\def\Tr{ {\,{\rm Tr}\,} }
\def\T{ {\rm T}\, }
\def\d{\partial}
\def\la{\raise.16ex\hbox{$\langle$} \, }
\def\ra{\, \raise.16ex\hbox{$\rangle$} }
\def\st{\, \raise.16ex\hbox{$|$} \, }
\def\go{\rightarrow}

\def\CS{ {\rm CS} }
\def\EM{ {\rm EM} }

\def\tot{ {\rm tot} }
\def\ext{ {\rm ext} }

\def\onehalf{ \hbox{${1\over 2}$} }

\def\psibar{ \psi \kern-.65em\raise.6em\hbox{$-$} }
\def\Lbar{ {\cal L} \kern-.65em\raise.6em\hbox{$-$} }
\def\Dbar{ D \kern-.70em\raise.55em\hbox{$-$} }
\def\Pibar{ \Pi \kern-.77em\raise.58em\hbox{$-$} }
\def\Dbark{ D_k \kern-.75em\raise.55em\hbox{$-$} }
\def\sDbar{ D_k \kern-.75em\raise.55em\hbox{$-${\sevenrm 2}} }
\def\rhobar{ {\bar\rho} }


\global\newcount\refno
\global\refno=1 \newwrite\reffile
\newwrite\refmac
\newlinechar=`\^^J
\def \ref#1#2{\the\refno\nref#1{#2}}
\def\nref#1#2{\xdef#1{\the\refno}%
 \ifnum\refno=1\immediate\openout\reffile=refs.tmp\fi%
 \immediate\write\reffile{\noexpand\item{\noexpand#1.}#2\noexpand\nobreak}
 \immediate\write\refmac{\def\noexpand#1{\the\refno}}%
 \global\advance\refno by1}


\def\semi{\hfil\noexpand\break ^^J}  

\def\refn#1#2{\nref#1{#2}}


\def\ap#1#2#3{{\nineit Ann.\ Phys.\ (N.Y.)} {\ninebf {#1}}, #3 (19{#2})}

\def\ijmpA#1#2#3{{\nineit Int.\ J.\ Mod.\ Phys.} {\ninebf {A#1}}, #3 (19{#2})}
\def\ijmpB#1#2#3{{\nineit Int.\ J.\ Mod.\ Phys.} {\ninebf {B#1}}, #3 (19{#2})}

\def\mplA#1#2#3{{\nineit Mod.\ Phys.\ Lett.} {\ninebf A{#1}}, #3 (19{#2})}

\def\plB#1#2#3{{\nineit Phys.\ Lett.} {\ninebf {#1}B}, #3 (19{#2})}

\def\np#1#2#3{{\nineit Nucl.\ Phys.} {\ninebf B{#1}}, #3 (19{#2})}
\def\prl#1#2#3{{\nineit Phys.\ Rev.\ Lett.} {\ninebf #1}, #3 (19{#2})}
\def\prB#1#2#3{{\nineit Phys.\ Rev.} {\ninebf B{#1}}, #3 (19{#2})}
\def\prD#1#2#3{{\nineit Phys.\ Rev.} {\ninebf D{#1}}, #3 (19{#2})}

\def\ibid#1#2#3{{\nineit ibid.} {\ninebf {#1}}, #3 (19{#2})}

\def\cline{\hfil\noexpand\break  ^^J}

\immediate\openout\refmac=refno.tex


\refn\Arovas{
D. Arovas, J. R. Schrieffer, F. Wilczek, and A. Zee, \np {251} {85} {117}.}

\refn\Goldhaber{
A. Goldhaber, R. MacKenzie, and F. Wilczek, \mplA {4} {89} {21};
\semi X.G. Wen and A. Zee, {\nineit J. Phys. France}
{\ninebf 50} 1623, (1989).}

\refn\JackiwPi{
R. Jackiw and S.Y. Pi, \prD {42} {90} {3500}. }

\refn\Ho{
C.-L. Ho and Y. Hosotani,  \ijmpA {7} {92} {5797}.}

\refn\CKLee{
C.\ Kim, C.\ Lee, P.\ Ko, B-H.\ Lee, and H.\ Min,
\prD {48} {93} {1821}.}


\refn\Jain{
J.\ Jain, \prl {63} {89} {199}; \prB {40} {89} {8079}.}

\refn\Lopetz{
A.\ Lopez and E.\ Fradkin, \prB {44} {91} {5246}.}

\refn\Halperin{
B.I.\ Halperin, P.A.\ Lee, and N.\ Read, \prB {47} {93} {7312}.}

\refn\Simon{
S.H.\ Simon and B.I.\ Halperin, \prB {48} {93} {17368}.}

\refn\LaughlinOne{
V. Kalmeyer and R. B. Laughlin, \prl{59}{87}{2095};
\semi R. B. Laughlin, {\nineit Science} {\ninebf 242}, 525 (1988);
\semi R. B. Laughlin, \prl{60}{88}{2677}.   }

\refn\FHL{
A. L. Fetter, C. B. Hanna, and R. B. Laughlin, \prB{39} {89} {9679}.}

\refn\HLF{
C. B. Hanna, R. B. Laughlin, and A. L. Fetter, \prB{40}{89}{8745};
\ibid {43} {91} {309}.}

\refn\CWWH{
Y.-H. Chen, F. Wilczek, E. Witten and B. Halperin, \ijmpB {3} {89} {1001}.  }

\refn\WenZee{
X. G. Wen and A. Zee, \prB{41} {90} {240}.  }

\refn\CanGirv{
G. S. Canright, S. M. Girvin, and A. Brass, \prl {63} {89} {2291, 2295}. }

\refn\HosoChak{
Y. Hosotani and S. Chakravarty, IAS report, IASSNS-HEP-89/31,
 May 1989 ; \prB{42}{90}{342}.}

\refn\Fradkin{
E. Fradkin, \prl{63} {89} {322}; \prB {42} {90} {570}. }

\refn\FisherLee{
M.P.A. Fisher and D.H. Lee, \prl{63} {89} {903}. }

\refn\Mori{
H. Mori, \prB {42} {90} {184}.}

\refn\Banks{
T. Banks and J. Lykken, \np {336} {90} {500}. }

\refn\RDSS{
S. Randjbar-Daemi, A. Salam, and J. Strathdee, \np{340} {90} {403}. }

\refn\KitaMura{
Y. Kitazawa and H. Murayama, \np{338} {90} {777}; \prB{41}{90}{11101}.}

\refn\Sakita{
P. K. Panigrahi, R. Ray, and B. Sakita, \prB {42} {90} {4036}. }

\refn\Lykken{
J.D. Lykken, J. Sonnenschein and N. Weiss, \prD {42} {90} {2161};
\ijmpA{6}{91}{1335} }

\refn\HHL{
J. E. Hetrick, Y. Hosotani, and B.-H. Lee, \ap{209} {91} {151}. }

\refn\HosotaniOne{
Y.\ Hosotani, \ijmpB {7} {93} {2219}.}

\refn\CSFracStat{
A.M. Polyakov, \mplA{3}{88}{325};
\semi M.\ L\"uscher, \np {326} {89} {557};
\semi S.\ Iso, C.\ Itoi, and H.\ Mukaida,  \np {346} {90} {293};
        \plB {236} {90} {287}.}

\refn\Deser{
R.\ Jackiw and S.\ Templeton, \prD {23} {81} {2291};
\semi J.\ Shonfeld, \np {185} {81} {157};
\semi S.\ Deser, R.\ Jackiw and S.\ Templeton, \prl {48} {82} {975};
   \ap {140} {82} {372}.}

\refn\HosotaniTwo{
Y.\ Hosotani, \plB {319} {93} {332}; Minnesota preprint UMN-TH-1239/94 (Feb.
94).}

\refn\antiferro{
A.\ Lopez, A.G.\ Rojo and E.\ Fradkin, Illinois preprint,
{\nineit`` Chern-Simons
theory of the anisotropic quantum Heisenberg antiferromagnet on a square
lattice''}.}

\refn\Feynman{
Useful relations  are (8-10) and (8-12),  with typos corrected,
in R.P.\ Feynman and
A.R.\ Hibbs, {\nineit ``Quantum Mechanics and Path Integrals''} (McGraw-Hill,
1965).}

\refn\FradkinTwo{
A.\ Lopez and e.\ Fradkin, \prB {44} {91} {5246}.}

\refn\Sumantra{
A similar formula has been derived in
S.\ Chakravarty, \np {390} {93} {691}.}

\refn\Abrikosov{
See e.g.\ A.A.\ Abrikosov, L.P.\ Gorkov, and I.E.\ Dzyaloshinski, {\nineit
``Methods of Quantum Field Theory in Statistical Physics''}
(Dover Publications, 1975).}

\refn\Dai{
See also
Q. Dai, J.L. Levy, A.L. Fetter, C.B. Hanna, and R.B. Laughlin,
\prB {46} {92} {5642}.}


\global\newcount\secno \global\secno=0
\global\newcount\appno \global\appno=0
\global\newcount\meqno \global\meqno=1
\global\newcount\figno \global\figno=1
\newwrite\eqmac
\def\eqn#1{
        \ifnum\secno>0
            \eqno(\the\secno.\the\meqno)\xdef#1{\the\secno.\the\meqno}%
            \immediate\write\eqmac{\def\noexpand#1{\the\secno.\the\meqno}}%
        \else\ifnum\appno>0
                  \eqno(\romappno.\the\meqno)\xdef#1{\romappno.\the\meqno}%

\immediate\write\eqmac{\def\noexpand#1{\romappno.\the\meqno}}%
               \else
                  \eqno(\the\meqno)\xdef#1{\the\meqno}%
            \immediate\write\eqmac{\def\noexpand#1{\the\meqno}}%
                    \fi
        \fi
        \global\advance\meqno by1
          }
\newwrite\figmac
\def\fig#1{\ifnum\secno>0
            \the\figno\xdef#1{\the\figno}%
        \fi
        \global\advance\figno by1
          }

\immediate\openout\eqmac=energy.eq

\def\mytitle{The Energy Density in the Maxwell-Chern-Simons Theory}
\def\myname{D.\ Wesolowski, Y.\ Hosotani and S.\ Chakravarty }

\def\firstheadline{\hbox to \pagewidth{%
    {\baselineskip=10pt
     {\ninerm \hbox{\vtop{ \hsize=2.8truecm \parindent=0pt
           \leftline{Preprint from}
           \leftline{University of Minnesota}     }}}
      \hfil
          {\ninerm  \hbox{\vtop{ \hsize=2.8truecm \parindent=0pt
                \rightline{UMN-TH-1246/94}
                \rightline{\mydate}          }}} }%
       }}

\def\otherheadline{\nineit
  \ifodd\pageno \qquad \hss \mytitle \hss Page \folio
    \else Page \folio\hss \myname \hss\qquad\fi}
\headline={\ifnum\pageno=1 \firstheadline \else\otherheadline\fi}



\baselineskip=9pt

\vglue 2.cm

\baselineskip=25pt

\centerline{\bfBig The Energy Density in the Maxwell-Chern-Simons Theory}

\vskip 1.5cm

\baselineskip=12pt

\centerline{\tenrm  Denne Wesolowski and Yutaka Hosotani}


\centerline{\eightit School of Physics and Astronomy, University of Minnesota}
\centerline{\eightit Minneapolis, MN 55455}

\centerline{\ninerm and }
\centerline{\tenrm Sumantra Chakravarty}

\centerline{\eightit Department of Mechanical Engineering, University of
        Minnesota}
\centerline{\eightit Minneapolis, MN 55455}

\vskip 0.5cm

\centerline{\eightsl Type-set by plain \TeX }

\vskip .5cm

\baselineskip=10pt
{\parindent=15pt
{\eightrm
\midinsert \narrower\narrower   
A two-dimensional  nonrelativistic fermion system coupled to both
electromagnetic gauge fields and Chern-Simons gauge fields is analysed.
Polarization tensors relevant in the quantum Hall effect and
anyon superconductivity are obtained as simple closed integrals
and are evaluated numerically for all momenta and frequencies.  The
correction to the energy density  is evaluated in the random phase
approximation
(RPA),  by summing an infinite series of ring diagrams.    It is found that the
correction  has  significant dependence on the particle number density.

In the context of anyon superconductivity, the  energy density relative to the
mean field value is minimized  at a hole concentration
per lattice plaquette (0.05{\smallsim}0.06){\smallcdot
}({\eightit p}$_{\sixmit c}${\eightit a}/{\smallhbar})$^{\fiverm 2}$
where {\eightit p}$_{\sixmit c}$ and {\eightit a} are the  momentum cutoff and
lattice constant, respectively.   At the minimum the correction
is about {\smallminus}5 \% {\smallsim} {\smallminus}25 \%, depending on
the ratio 2{\eightit m}{\smallomega}$_{\sixmit c}$/{\eightit p}$_{\sixmit
c}^{\fiverm 2}$ where {\smallomega}$_{\sixmit c}$ is the frequency cutoff.

In the Jain-Fradkin-Lopez picture of the fractional quantum Hall effect
the RPA correction to the energy density is very large.  It diverges
logarithmically as the cutoff is removed, implying that corrections beyond
RPA become important at large momentum and frequency.
\endinsert
}
}



\vskip .6cm
\normal

\secno=1  \meqno=1

\leftline{\bf 1. Introduction}

There is a growing interest in Chern-Simons theory.   Gauge fields with a
pure Chern-Simons term, which we call Chern-Simons gauge fields,
effectively alter the statistics of matter fields.\myref{\Arovas-\CKLee}
In Jain's picture of the fractional quantum Hall effect,\myreff{\Jain}
 an electron and  magnetic flux form a
bound state, or a composite fermion.  The fractional quantum Hall effect
for electrons is understood as integral quantum Hall effect for these
composite fermions.  Further, Fradkin and Lopez have shown that these
composite fermions are regarded as fermions interacting with  Chern-Simons
gauge fields.\myref{\Lopetz}   Halperin, Lee, and Read have advocated this
picture further to  investigate physics near half filling in the quantum Hall
effect.\myref{\Halperin,\Simon}

In anyon superconductivity, physics of essentially the same system of
fermions coupled to Chern-Simons and Maxwell gauge fields is explored, but with
a  different Chern-Simons coefficient.\myref{\LaughlinOne-\HosotaniOne}  It is
known
that  this system shows a rather unique behavior at finite temperature,
particularly around $T \sim 100$K.\myref{\HHL,\HosotaniOne}  So far there has
been no evidence for  its possible connection to high $T_c$ superconductors.
In fact, the theory of anyon superconductivity is still in its infancy to the
extent that  it provides only few  predictions which can be tested
experimentally.

In the relativistic case the Chern-Simons gauge fields induce
boson-fermion transmutation.\myref{\CSFracStat}  If gauge fields have both
Maxwell and Chern-Simons terms, the Chern-Simons term gives the gauge fields a
topological mass.\myref{\Deser} Further it has been recently shown that the
presence of Dirac fermions leads, under certain conditions, to the spontaneous
breakdown of the Lorentz invariance by dynamical generation of a magnetic
field ($B\not= 0$).\myref{\HosotaniTwo}  There  quantum fluctuations play a
crucial role in lowering the energy density of the true ground state with
$B\not= 0$.
Lopez, Rojo, and Fradkin have shown that an effective relativistic  Dirac
theory with Maxwell-Chern-Simons gauge interactions naturally arises in the
quantum Heisenberg antiferromagnet on a square lattice.\myref{\antiferro}

In this paper we examine a nonrelativistic fermion system
interacting with both Maxwell and Chern-Simons gauge fields.
Employing the technique developed in ref.\ [\HosotaniTwo],  we first cast
one-loop polarization tensors for gauge fields  in simple integral forms
suited for both analytical and numerical evaluation.  Then we derive the
exact formula for the energy density in terms of gauge field propagators.
Applying  our one-loop formula for the polarization tensors, which
corresponds to the random phase approximation (RPA), or to
summing up an infinite series of ring diagrams in perturbation theory, we
examine contributions of quantum fluctuations to the energy density.

The detailed numerical evaluation is given in the absence of dynamical
electromagnetic interactions.   We shall find that RPA corrections generate
non-trivial dependence of the energy density on the particle (or hole) number
density in the context of anyon superconductivity.  The energy is minimized
at a number density precisely where high $T_c$ material is superconducting.
In the case of the fractional quantum Hall effect, RPA gives a large
correction to the energy density, in accord with the equivalence between
the electron picture and the composite fermion picture.

\vskip .5cm

\secno=2 \meqno=1

\leftline{\bf 2. The model}

The model we consider consists of a nonrelativistic fermion field $\psi$,
electromagnetic field $A_\mu$, and Chern-Simons gauge field $a_\mu$.  Its
Lagrangian is given by
$$\eqalign{
\L =  {1 \over 2} \,( \epsilon E_j^2 -\chi B^2 )
      &- {\N \over 2} \, \eps a_\mu \d_\nu a_\rho
 + i\psi^\dagger D_0 \psi - {1\over2m} (D_k\psi)^\dagger (D_k\psi)
  + e\rhobar A_0 \cr
&\hskip 0cm E_j= F_{0j} ~~,~~ B = - F_{12}= \d_1 A^2 - \d_2 A^1  \cr
&\hskip 0cm D_\mu = \d_\mu  + i(a_\mu + eA_\mu) ~~~.  \cr}  \eqn\model $$
The last term represents a contribution from a background neutralizing
charge.  Note that $a^k=-a_k$ and $A^k=-A_k$.  The Euler equations are
$$\eqalign{
&\epsilon \, \d_k E_k= e \, (j^0 - \rhobar) \cr
&\ep^{kl} \, \chi \, \d_l B \, =  e \,  j^k \cr
&- {\N\over 2} \, \eps f_{\nu\rho} = ~ j^\mu  \cr
&i\d_0 \psi = \Big\{  -{1\over 2m} \, D_k^2 + a_0 + eA_0  \Big\} \psi  \cr}
       \eqn\EulerEqs  $$
where the currents are given by
$$\eqalign{
j^0 &= \psi^\dagger \psi \cr
j^k & = - {i\over 2m} \, \big\{ \psi^\dagger D_k \psi
    - (D_k \psi)^\dagger \psi \big\} ~~~.  \cr}  \eqn\current  $$

Suppose a uniform external magnetic field $B_\ext$ is applied  and the
system remains translation invariant.  Eq.\ (\EulerEqs) implies that $\psi$
feels a total magnetic field
$$b_\tot \equiv  b^{(0)} + eB^{(0)}
= {\rhobar\over \N} + e B_\ext  ~~~.  \eqn\tatalB  $$
The Landau level densities associated with the external magnetic field
and the total magnetic field are given by
$$\eqalign{
\rho_L^\ext &= {|eB_\ext|\over 2\pi}    \cr
\rho_L^\tot &= {|b_\tot|\over 2\pi}   ~~~. \cr} \eqn\Ldensity $$
Something special happens when the total filling factor is an integer:
$$\nu = {\rhobar\over \rho_L^\tot}= n ~~~.  \eqn\filling   $$
The change in the statistics phase induced by Chern-Simons gauge fields is
$\Delta \theta_s =1/2\N$.  Combining (\tatalB), (\Ldensity), and (\filling),
one finds
$${\Delta \theta_s\over \pi} = {1\over 2\pi\N} =
\ep(b_\tot) \, {1\over n} - \ep(eB_\ext) \, {1\over \nu_\ext}
  ~~~,~~~ \nu_\ext = {\rhobar\over \rho_L^\ext} ~~~.   \eqn\specialCS  $$

In the Jain-Fradkin-Lopez picture, the system exhibits the fractional
quantum Hall effect when $\Delta \theta_s$ is a multiple of $2\pi$.
Suppose that $eB_\ext >0$.  Then the condition is satisfied if
$$\eqalign{
{1\over 2\pi \kappa} &= - 2 p  \cr
\nu_\ext &= {n\over 2pn \pm 1}  \cr}
     \eqn\FQHEvalue $$
where $\pm$ corresponds to $\ep(b_\tot)$.
In the case $eB_\ext <0$ the signs of $\kappa$ and $b_\tot$ are reversed.
The main sequence in fractional quantum Hall effect is given by
$p=1$, or $\nu_\ext=n/(2n \pm 1)$.
In this paper we focus on the special case $\nu=1$.  Generalization to
the case $\nu =n$ is straightforward, but is left for a future investigation.
Note that for $\nu=1$
$$\bigg\{ {1\over 2\pi \kappa} - \ep(b_\tot) \bigg\} \, \rhobar
 = - {eB_\ext \over 2\pi} ~~.   \eqn\fillingOne $$

\vskip .5cm

\secno=3 \meqno=1

\leftline{\bf 3.  Fermion propagator}

The perturbation theory is defined around $b_\tot$.    We rewrite
$a_\mu \go a_\mu^{(0)} + a_\mu$ and $A_\mu \go A_\mu^\ext +A_\mu$,
i.e.\ $a_\mu$ and $A_\mu$ now represent fluctuation parts.  Then
the Lagrangian becomes
$$\L ~ = -{\chi\over 2} B_\ext^2  +\L_0^{\rm gauge}[a,A]
   + \L_0^{\rm matter}[\psi,\psi^\dagger] + \L_{\rm int}[a,A,\psi,\psi^\dagger]
                 ~~.  \eqn\LagrangianTwo  $$
$\L_0^{\rm gauge}+ \L_0^{\rm matter}$ defines the zeroth order ``free'' part.
The gauge field part is given by
$$\L_0^{\rm gauge} =  {1 \over 2} \,( \epsilon E_j^2 -\chi B^2 )
      - {\N \over 2} \, \eps a_\mu \d_\nu a_\rho  + \L_{\rm g.f.}[a,A]
        \eqn\LagrangianThree  $$
where $\L_{\rm g.f.}[a,A]$ is a gauge fixing term to be specified in the
next section.
The matter field part is
$$\eqalign{
\L_0^{\rm matter}&=  i\psi^\dagger \d_0 \psi
   - {1\over2m} (\Dbar_k\psi)^\dagger (\Dbar_k\psi) \cr
 \Dbar_k &= \d_k - i (a^{(0)k} + eA_\ext^k) ~~~.   \cr}
    \eqn\LagrangianFour  $$

The interaction part is given by
$$\eqalign{
\L_{\rm int} =&  - \chi B_\ext B - (\psi^\dagger \psi - \rhobar) \, (a_0+eA_0)
            \cr
&- (a^k + eA^k) {i\over 2m} \Big\{
   \psi^\dagger (\Dbar_k\psi) - (\Dbar_k \psi)^\dagger \psi \Big\}
 - {1\over 2m}  (a^k+ eA^k)^2 \,\psi^\dagger \psi ~~. \cr}
    \eqn\LagrangianFive  $$
The $B_\ext B$ term in $\L_{\rm int}$,
being a total derivative, does not contribute in perturbation theory  and
may be dropped.

The zeroth order fermion propagator has been well described in the literature.
Usually it is given in the form of an infinite sum over the Landau level
index.  Our task here is to recast it in an integral form
for later convenience.

Let us quote the result from Ref.\ [\HosotaniOne].  In the Landau gauge
$$\eqalign{
a^{(0)k}+ eA_\ext^k &= - \ep(b_\tot) \, \delta^{k1} \, {x_2\over l^2} \cr
{1\over l^2} &= \big| b_\tot \big|  ~~~.  \cr}
   \eqn\LandauGauge  $$
$l$ is the magnetic length.
When $\nu$ lowest Landau levels are completely filled, the fermion
propagator is given by
$$\eqalign{
G(x,y) &= -i\la T[\psi(x)\psi^\dagger (y)]\ra    \cr
&= e^{- i \ep (b_\tot)  (x_1-y_1)(x_2+y_2) /2l^2 } \cdot G_0(x-y)  \cr
G_0(x) &= -i\bigg\{ \theta(x_0) \sum_{n=\nu}^\infty
  - \theta(-x_0) \sum_{n=0}^{\nu -1} \bigg\} \, e^{-i\ep_n x_0} \cr
&\hskip .5cm \times {1\over 2\pi l^2} \int_{-\infty}^\infty  dz\,
 e^{-izx_1 / l} ~v_n[z-\bz(x_2)] ~ v_n[z+ \bz(x_2)]   \cr}
   \eqn\FpropagatorOne    $$
where
$$\ep_n = \Big( n+ {1\over 2} \Big) {1\over ml^2}
   \qquad (n=0,1,2,\cdots)        \eqn\LandauLevel $$
is the $n$-th Landau energy level, and
$$\eqalign{
v_n(x) & ={(-1)^n \over 2^{n/2} \pi^{1/4} (n!)^{1/2}}
       e^{x^2/ 2} {d^n\over dx^n} e^{-x^2}    \cr
\bz(x_2) &= \ep(b_\tot) \, {x_2\over 2l}   ~~~. \cr}
          \eqn\eigenfunctions   $$
The phase factor in the propagator $G(x,y)$ is not translation invariant,
but does not contribute to any physical quantities as we see below.

We now utilize the Feynman-Hibbs summation formula\myref{\Feynman}
$$\sum_{n=0}^\infty e^{-int} \, v_n(x)v_n(y)
= {e^{it/2}\over \sqrt{2\pi i\sin t} } ~
\exp \bigg\{ {i\over 2 \sin t} \Big[ (x^2+y^2) \cos t - 2xy \Big] \bigg\}
      \eqn\FHformulaOne   $$
in order to perform the infinite sum and obtain an explicit integral
expression for the propagator.  We also quote another
formula
$$\sum_{n=0}^\infty  {v_n(x)v_n(y) \over n-\lambda - i\ep}
= i \int_0^\infty dt ~
{e^{i(\lambda +{1\over 2})t - \ep t}\over \sqrt{2\pi i\sin t} } ~
\exp \bigg\{ {i\over 2 \sin t} \Big[ (x^2+y^2) \cos t - 2xy \Big] \bigg\}
      \eqn\FHformulaTwo   $$
which follows from (\FHformulaOne).

First we evaluate the propagator in the $\nu=0$ case.  Employing
(\FHformulaOne)
and taking the Fourier transform
$$G_0(p) = G_0(\omega ,\vp) = \int dx  ~
G_0(x) \,  e^{i(\omega x_0-\vp \vx )}~,           $$
one finds that
$$G_0(p) \big|_{\nu=0}
= -i ml^2 \int_0^\infty {dt \over \cos (t/2) } ~
  e^{-i(p_1^2+p_2^2) l^2 \tan (t/2)  +i\omega ml^2t}     ~~~.
      \eqn\FpropagatorTwo$$
For $\nu \ge 1$ we observe that
$$\eqalign{
&G_0(x)\big|_\nu -G_0(x) \big|_{\nu=0}  \cr
&= \sum_{n=0}^{\nu-1} e^{-i\ep_n x_0}\, {i\over 2\pi l^2}
 \int_{-\infty}^\infty dz ~ e^{-izx_1/l}
 \, v_n[z - \bz (x_2)] v_n[z + \bz (x_2)]  ~~.  \cr}
         \eqn\FpropagatorThree   $$
Using the explicit expression for $v_n$, one finds that for $\nu=1$, for
instance,
$$\eqalign{
G_0(p) \big|_{\nu=1}
=& 4\pi i \,  e^{-\vp^2l^2} \, \delta (\omega -\ep_0) \cr
& -iml^2 \int_0^\infty {dt \over \cos (t/2) } ~
  e^{-i\vp^2 l^2 \tan (t/2)  +i\omega ml^2t}~~~.  \cr}
     \eqn\FpropagatorFour  $$

\vskip .5cm

\secno=4  \meqno=1

\leftline{\bf 4.  Gauge field propagators}

Free gauge field propagators are obtained from $\L_0^{\rm gauge}$ in
(\LagrangianThree).   For the sake of unified treatment of both
electromagnetic and Chern-Simons gauge fields, let us consider gauge
fields described by
$$\Lag_0 =  {1 \over 2} \,( \epsilon E_j^2 -\chi B^2 )
      - {\N \over 2} \, \eps A_\mu \d_\nu A_\rho
   -{1\over 2 \chi} \, (\epsilon \d_0 A^0 + \chi \d_kA^k)^2 ~~.
   \eqn\generalGaugeField   $$
The last term represents gauge-fixing.  Electromagnetic fields
are described with $\N=0$, whereas Chern-Simons fields in the
$\epsilon, \chi \go 0$ limit:
$$\L_0^{\rm gauge}[a,A]
 =  \Lag_0[A; \N=0] + \Lag_0[a; \epsilon, \chi \go 0] ~~.
            \eqn\gaugeLagrangian   $$
As is obvious, the propagator is singular in the $\epsilon, \chi \go 0$
limit, which is merely a gauge artifact.  Physical quantities
such as the energy density are well defined in the $\epsilon, \chi \go 0$
limit.  We take this limit when convenient and safe to do so.

In passing, we mention that for pure Chern-Simons gauge fields the alternative
gauge fixing term
$$\L_{\rm g.f.} =   \, {1\over 2\alpha} \, (\d_k a^k)^2
   \eqn\GaugeFixing  $$
is also very useful and convenient.\myref{\HosotaniOne}  The radiation gauge
is obtained in  the $\alpha \go 0$ limit.  We adopt the gauge fixing in
(\generalGaugeField) in this paper.

The Lagrangian (\generalGaugeField) has the form
$$\eqalign{
\Lag_0&={1\over 2} \, A_\mu K^{\mu\nu} A_\nu  ~~~, \cr
K^{\mu\nu} &=  \left( \matrix{
    (\epsilon/\chi) \D  & \kappa \partial_2 & -\kappa \partial_1 \cr
    -\kappa \partial_2 & -\D & \kappa \partial_0 \cr
    \kappa \partial_1 & -\kappa \partial_0 & -\D \cr}  \right) ~~~, \cr
\D&= \epsilon \partial_0^2 -  \chi \nabla^2 ~~~.   \cr}
            \eqn\generalLagrangian $$
For this Lagrangian, the propagator, the inverse of $K^{\mu\nu}$,
 is straightforwardly found to be
$$(D_0)_{\mu \nu} = {-1\over \epsilon \D +\kappa^2}
\left\{  \pmatrix{\chi &0 &0 \cr
                  0& -\epsilon &0 \cr
                  0& 0& -\epsilon \cr}
+ \chi \kappa^2 \, {\partial_\mu \partial_\nu\over \D^2}
+ {\kappa\over \D}
  \pmatrix {0& \chi\partial_2& -\chi\partial_1 \cr
       -\chi\partial_2 & 0& \epsilon\partial_0 \cr
       \chi\partial_1 & \epsilon\partial_0 &0 \cr} \right\}  ~~.
             \eqn\gaugePropagator  $$
Equal time commutation relations are given by
$$\eqalign{
&[A_\mu(t,\x) , \Pi^\nu(t,\y) ] = i {\delta_\mu}^\nu \, \delta(\x-\y) \cr
&[A_\mu(t,\x) , A_\nu(t,\y) ] = [\Pi^\mu(t,\x) , \Pi^\nu(t,\y) ] =0 \cr
&\Pi^0  = - {\epsilon^2\over \chi} \d_0 A_0  \cr
&\Pi^j  = \epsilon \d_0 A_j - {\N\over 2} \ep^{jk} A_k   \cr}
  \eqn\commutator  $$

In evaluating the energy density we need the full gauge field propagators.
The interaction, given by (\LagrangianFive), mixes the electromagnetic
and Chern-Simons gauge fields.   The propagators have a matrix structure:
$$\hat D_{\mu\nu} = -i \left( \matrix{
     \la \T [A_\mu A_\nu]\ra  & \la \T [A_\mu a_\nu] \ra \cr
     \la \T [a_\mu A_\nu]\ra  & \la \T [a_\mu a_\nu] \ra \cr}  \right) ~~.
         \eqn\FullPropagatorOne  $$
When the lowest Landau level is completely filled so that the perturbative
ground state  is of Hartree-Fock type, one can develop a Feynman diagram
method.   Since both gauge fields couple to fermions in the combination of
$a_\mu + eA_\mu$, one can write
$$\eqalign{
{\hat D}^{-1}_{\mu\nu}
&= \hat {D_0}^{-1}_{\mu\nu} - \hat \Gamma_{\mu\nu} \cr
&= \left( \matrix{
  D_0^\EM{}^{-1}_{\mu\nu} & 0 \cr
  0 &  D_0^\CS{}^{-1}_{\mu\nu} \cr}  \right)
 - \Gamma_{\mu\nu} \left( \matrix{ e^2 & e \cr e & 1 \cr} \right) ~~. \cr}
         \eqn\FullPropagatorTwo  $$
Here $\Gamma^{\mu\nu}$ represents the sum of one-particle irreducible diagrams
common to both $a_\mu$ and $e A_\mu$.

Gauge invariance, translational invariance, and rotational invariance imply
that $\Gamma^{\mu\nu}$ is expressed in terms of three independent invariant
functions $\Pi_k$'s:
$$\eqalign{
\Gamma^{\mu\nu}(p)
=&  ~ (p^\mu p^\nu - p^2 g^{\mu\nu} ) \Pi_0
 + i \ep^{\mu\nu\rho} p_\rho \Pi_1 \cr
& +(1-\delta^{\mu 0} ) (1-\delta^{\nu 0})
  (p^\mu p^\nu - {\vec p\,}^2 \delta^{\mu\nu}) (\Pi_2 - \Pi_0) . \cr}
  \eqn\GammaDecomposition  $$
All $\Pi_k$'s are functions of $p_0^2$ and $\vp^2$.

\vskip 1cm

\secno=5   \meqno=1

\leftline{\bf 5.  Evaluation of {\greekPi}$_{\bf k}$'s}

We evaluate the kernel $\Gamma^{\mu \nu} (\omega ,\vq )$,
or equivalently the invariant functions $\Pi_k$'s, to the leading order
in the case $\nu=1$.  The interaction is given by (\LagrangianFive).  One-loop
contributions are  depicted in Fig.\ 1.

To calculate $\Pi_k$'s we take a frame $\vq=(q,0)$.  Then
$$\eqalign{
q^2\Pi_0 &= \Gamma^{00} = \Gamma^{(b)00}  \cr
iq\Pi_1 &= \Gamma^{02}  = \Gamma^{(c)02} \cr
 \omega^2 \Pi_0 - q^2 \Pi_2
&= \Gamma^{22}= \Gamma^{(a)22} + \Gamma^{(d)22} \equiv \Pibar_2 ~~ . \cr}
        \eqn\PiFormulaOne  $$
Diagram (a) is easily evaluated to give
$$\Gamma^{(a)jk}(\omega , \vq ) = -{\bar\rho \over m}\delta^{jk}.$$
As is well known, other diagrams yield
$$\eqalign{
\Gamma^{(b)00}(q) &= i\int {d^3p\over {(2\pi)^3}} \, G_0(p)G_0(p-q)  \cr
\Gamma^{(c)0j}(q) &= -{1\over 2m} \int {d^3p\over{(2\pi)^3}} \,
  \Big\{ G_0(p)\cdot D_j^-G_0(p-q)
  + D_j^+G_0(p)\cdot G_0(p-q) \Big\}    \cr
\Gamma^{(d)22}(q) &= {-i\over{4m^2}}\int {d^3p\over (2\pi )^3} \,
  \Big\{
D_2^+G_0(p)\cdot  D_2^-G_0(p-q) + D_2^+G_0(p)\cdot D_2^+G_0(p-q)  \cr
&\hskip 2.2cm  + D_2^+D_2^-G_0(p)\cdot G_0(p-q)
+ G_0(p)\cdot D_2^-D_2^+G_0(p-q) \Big\}   \cr
&D_2^\pm G_0(p) = \Big( ip_2 \pm
\epsilon (b_\tot) {1\over 2l^2}
{\d\over \d p_1} \Big) \, G_0(p) ~~~.  \cr}
          \eqn\PiFormulaTwo $$
The phase factor
in $G(x,y)$  (\FpropagatorOne), does not contribute to
$\Gamma^{\mu\nu}$.

The next step is to insert (\FpropagatorFour) into (\PiFormulaTwo) and perform
the momentum integrals.    All integrals
involve heavy, but similar  manipulations.  We shall describe the computation
of  $\Pi_0$ in detail.

There are four terms in the integral
$$\eqalign{
&q^2 \Pi_0(q) = i\int {d^3p\over {(2\pi)^3}}  \cr
&\times \Big\{ 4\pi i \,  e^{-\vp^2l^2} \, \delta (p_0 -\ep_0)
 -iml^2 \int_0^\infty {dt \over \cos \onehalf t } ~
  e^{-i\vp^2 l^2 \tan (t/2)  +ip_0 ml^2t}  \Big\}  \cr
&\times \Big\{ 4\pi i \,  e^{-(\vec p-\vq)^2l^2} \, \delta (p_0 -\omega -\ep_0)
 -iml^2 \int_0^\infty {dt' \over \cos \onehalf t' } ~
  e^{-i(\vec p-\vq)^2 l^2 \tan (t'/2)  +i(p_0-\omega) ml^2t'}  \Big\} ~. \cr}
     \eqn\IntegralOne  $$
The product of the two $\delta$-function terms is easily evaluated to be
$$- {i\over l^2} \, e^{-q^2l^2/2} \, \delta(\omega) ~~.  \eqn\subIntegralOne $$
Two cross terms of the $\delta$-function and $t$-integral pieces are, after
$p_0$ integration,
$$\eqalign{
&4\pi i ml^2 \int {d\vp\over (2\pi)^3} \,
e^{-\vp^2l^2}  \int_0^\infty {dt \over \cos \onehalf t} ~
  e^{-i(\vec p-\vq)^2 l^2 \tan (t/2)  +i(\ep_0-\omega) ml^2t}   \cr
& \hskip 2cm + \big\{ (\omega, \vq) \go (-\omega, -\vq) \big\}   \cr
\noalign{\kern 5pt}
&=i {m\over 2\pi} \int_0^\infty {dt \over \cos \onehalf t} \,
{1\over 1 + i \tan \onehalf t}
  \,\exp \bigg( i(\ep_0 - \omega) ml^2 t
 - {i  q^2l^2\tan \onehalf t \over 1 +  i \tan \onehalf t}  \bigg) \cr
&\hskip 2cm + \big\{ (\omega, \vq) \go (-\omega, -\vq) \big\}   \cr
\noalign{\kern 5pt}
&=i {m\over \pi} \int_0^\infty dt \,
  e^{-  q^2l^2(1 - e^{-it})/2  } \, \cos(\omega ml^2 t)  ~~~.  \cr}
         \eqn\subIntegralTwo $$
The product of the two $t$-integrals in (\IntegralOne) vanishes:
$$\eqalign{
-&(ml^2)^2 \int_0^\infty {dt \over \cos \onehalf t}
  \int_0^\infty {dt' \over \cos \onehalf t'}  \int {d^3p\over (2\pi)^3} \cr
&\hskip .5cm \times
e^{- i\vp^2 l^2 \tan \onehalf t + i p_0 ml^2 t } \cdot
e^{- i(\vec p - \vq)^2 l^2 \tan \onehalf t' + i (p_0 - \omega) ml^2 t' } \cr
&= \int_0^\infty dt  dt' \int d\vp ~\cdots  \delta(t+t') \cdots  \cr
&=0 ~~~.  \cr}   \eqn\subIntegralThree  $$
The manipulation in the last equality is justified since the integrand
is regular at $t=t'=0$.

$\Pi_0(q)$ is a regular function of $\omega$ so that the $\delta(\omega)$
term in (\subIntegralOne) must be cancelled by a part of (\subIntegralTwo).
To see this we rewrite (\subIntegralTwo) as
$$\eqalign{
&i {m\over 2\pi} \int_0^\infty dt \,
  e^{- q^2l^2(1 - e^{-it})/2  } \,
  \big\{ e^{i(\omega+ i\ep)  ml^2 t} + e^{ -i(\omega- i\ep)  ml^2 t}  \big\}
\cr
&=  {m\over 2\pi} \int_0^\infty dt \,
  e^{- q^2l^2(1 - e^{-it})/2  } \, {\d\over \d t}
  \bigg\{ {e^{i(\omega+ i\ep)  ml^2 t} \over(\omega + i\ep) ml^2}
- {e^{ -i(\omega -i\ep)  ml^2 t}\over (\omega - i\ep) ml^2} \bigg\} ~~.\cr} $$
Integration by parts yields
$$\eqalign{
&{-1\over 2\pi l^2} \Big({1\over \omega + i\ep} - {1\over \omega - i\ep}\Big)
\cr
&\hskip .5cm
-  {m\over 2\pi} \int_0^\infty dt \,
 \Big( - i {q^2l^2\over 2} \, e^{-it} \Big) \, e^{- q^2l^2(1 - e^{-it})/2  }
 \bigg\{ {e^{i(\omega+ i\ep)  ml^2 t} \over (\omega + i\ep) ml^2}
- {e^{ -i(\omega - i\ep)  ml^2 t}\over (\omega - i\ep) ml^2} \bigg\} \cr
&= {i\over l^2} \delta(\omega)
+ {iq^2\over 4\pi} \int_0^\infty dt \, e^{-it - q^2l^2(1 - e^{-it})/2 }
\bigg\{ -2\pi i \delta(\omega) + {2i\over \omega} \sin (\omega ml^2 t) \bigg\}
          ~.  \cr}
        \eqn\subIntegralFour $$
Now the second $\delta(\omega)$ piece is
$$\eqalign{
 { q^2\over 2}\delta(\omega)
\int_0^\infty dt \, e^{-it - \ep t - q^2l^2(1 - e^{-it})/2 }
&=  { q^2\over 2}\delta(\omega)
\int_0^\infty (-ids) \, e^{-s  - q^2l^2(1 - e^{-s})/2 } \cr
&= {i\over l^2} ( e^{-q^2l^2/2} - 1 ) \, \delta(\omega)  ~~. \cr}
     \eqn\subIntegralFive  $$
Hence (\subIntegralTwo) becomes
$${i\over l^2} \, e^{-q^2l^2/2} \, \delta(\omega)
- {q^2 \over 2\pi \omega}
\int_0^\infty dt \, e^{-it - q^2l^2(1 - e^{-it})/2 }  \sin (\omega ml^2 t) ~~.
       \eqn\subIntegralSix   $$
The first term exactly cancels (\subIntegralOne).
Combining (\subIntegralOne), (\subIntegralThree), and (\subIntegralSix), one
finds that
$$\eqalign{
&\Pi_0 = {ml^2 \over 2\pi}\,F_0 \big( \onehalf q^2l^2 , \omega ml^2 \big) \cr
\noalign{\kern 5pt}
&F_0(x,y) = -{1\over y} \int_0^\infty dt ~ \sin y t \,
e^{-it - x(1 - e^{-it}) }      ~~~.  \cr}
    \eqn\PiZero   $$

The evaluation of $\Pi_1$ and $\Pibar_2$ proceeds similarly.   The cancellation
of $\delta$-function pieces  takes place as in the case of $\Pi_0$.
The result is
$$\eqalign{
&\Pi_1 = { \ep(b_\tot)\over 2\pi} \, F_1 \big( \onehalf q^2l^2 , \omega ml^2
                  \big) \cr
&\Pibar_2 =  {1\over 2\pi ml^2} \, F_2 \big( \onehalf q^2l^2 , \omega ml^2
                   \big) \cr
&F_1 (x,y) =  {\d\over \d x} x \cdot F_0(x,y) \cr
&F_2 (x,y) = -1 + \bigg\{ x +\Big({\d\over \d x} x \Big)^2 \bigg\} F_0(x,y)
{}~~~. \cr}
    \eqn\PiOneTwo  $$

In refs. [\FHL,\CWWH,\RDSS,\HosotaniOne,\FradkinTwo] these $\Pi_k$'s are
obtained in the form of infinite sums over the Landau level index $n$.
Applying the second Feynman-Hibbs formula (\FHformulaTwo) to the result
in ref.\ [\HosotaniOne],  the
results (\PiZero) and (\PiOneTwo) are reproduced with some labor.

Notice that all $\Pi_k$'s are related to one function $F_0$.
For small $x$ \myref{\Sumantra}
$$F_0(x,y) =  e^{-x} \sum_{k=0}^\infty {x^k\over k!}
   {-1\over y^2 - (k+1)^2 + i \ep}   ~~~.  \eqn\FzeroOne  $$
In particular
$$\eqalign{
F_0(0,y) &= {1\over 1 - y^2 - i\ep} \cr
F_0(x,0) &= {e^{-x}\over x} \int_0^x dw \, {e^w -1\over w} ~~. \cr}
   \eqn\FzeroTwo  $$

One can derive an alternative integral representation for $F_j$ which is more
suited for both analytical and numerical evaluation.  First we Wick-rotate the
$y$ variable: $\hat F_j(x,z) = F_j(x,iz)$.  Then
$$\eqalign{
\hat F_0(x,z) &= {e^{-x} \over x} \cdot  h(x,z) \cr
h(x,z) &= \sum_{k=0}^\infty {x^{k+1} \over k!} {1\over z^2 + (k+1)^2}   ~~.
\cr}
              \eqn\FzeroThree  $$
$h(x,z)$ satisfies
$$\eqalign{
\bigg\{ \Big( x {\d\over \d x} \Big)^2 + z^2 \bigg\} \, h(x,z) &= x e^x \cr
h(0,z) &= 0 \cr
\d_x h(0,z)  &= {1\over z^2 + 1}  ~~. \cr}
    \eqn\hOne   $$
In terms of $s= \ln x$,
$h(x,z)=z^{-1} \int_{-\infty}^s du \,\rho(u) \sin z(s-u)$ where
$\rho(s) = xe^x$.    Or
$$\eqalign{
h(x,z) &= {1\over z} \int_0^x dw \, \sin \Big( z \ln {x\over w} \Big) \cdot
     e^w  \cr
&= -{x\over z} \int_0^1 dt \, \sin (z\ln t) \cdot e^{xt}  ~~. \cr}
       \eqn\hTwo $$
This $h(x,z)$ satisfies the boundary condition in (\hOne).

Hence one finds that
$$\eqalign{
\hat F_0(x,z) &= - {1\over z} \int_0^1 dt \, \sin(z\ln t) \cdot e^{-(1-t)x} \cr
\hat F_1(x,z) &= - {1\over z} \int_0^1 dt \, \big\{ 1 +x(t-1) \big\} \,
\sin(z\ln t) \cdot e^{-(1-t)x} \cr
1+ \hat F_2(x,z) &= - {1\over z} \int_0^1 dt \,
  \big\{ 1 - 2x + 3xt + x^2(t-1)^2 \big\} \,
   \sin(z\ln t) \cdot e^{-(1-t)x}  ~.\cr}
          \eqn\FFormula  $$
The behavior of these functions is depicted in Fig.\ 2.  The asymptotic
behavior is given  by
$$\eqalign{
\hat F_0 &\sim {1\over x^2 + z^2 } \cr
\hat F_1 &\sim {z^2 - x^2 \over (x^2 + z^2)^2 } \cr
1 + \hat F_2 &\sim  {x\over x^2 + z^2 } + {1\over x^2 + z^2}
   - {8x^2 z^2 \over (x^2 + z^2)^3} ~~.  \cr}
     \eqn\asymptoticF   $$
These approximate expressions are valid to the accuracy of 2 \%   for
 $x$ or $z>10$.

\vskip .5cm

\secno=6 \meqno=1

\leftline{\bf 6.  Energy density}

Quantum fluctuations shift
the energy density of the ground state  from the mean field
value.  This shift can be related to the full
gauge field propagators.  We first derive an exact formula for the energy
density\myreff{\Abrikosov}, and estimate it by utilizing the result in the
previous section.   The mean field energy density for $\nu=1$ is given by
$$
\E_{\rm mean} = {1\over 2} B_\ext^2
+ {\pi \rhobar^2\over m}   ~~.  \eqn\meanEnergy $$
Here $B_\ext$ is related to $\rhobar$ and $\kappa$ by  (\fillingOne).
The interactions  (\LagrangianFive)
give corrections to  $\E_{\rm mean}$.   The corresponding interaction
Hamiltonian is
$$\eqalign{
H_{\rm int} &= H^{(1)} + H^{(2)}  \cr
H^{(1)}&= \int d\x \, \Big\{
(a^k + eA^k) {i\over 2m} \Big(
   \psi^\dagger \Dbar_k\psi - (\Dbar_k \psi)^\dagger \psi \Big)
 + (a_0 + eA_0) (\psi^\dagger \psi - \rhobar) \Big\}  \cr
H^{(2)}&= \int d\x \, {1\over 2m} (a^k+ eA^k)^2 \,\psi^\dagger \psi ~~.  \cr}
       \eqn\HInteractionOne   $$

Now we consider a Hamiltonian given by
$$H(g) = H_0 + g H^{(1)} + g^2 H^{(2)}   \eqn\TotalH   $$
where $H_0$ is the free part of the Hamiltonian.
With an auxiliary parameter $g$, $H(g)$ connects the
mean field Hamiltonian $H(0)$ and the full Hamiltonian $H(1)$.  Its ground
state satisfies
$$\eqalign{
H(g) | \Psi (g)\ra &=E(g) |\Psi (g)\ra \cr
E(g) &=\la \Psi (g) | H(g) | \Psi (g)\ra ~~. \cr}  \eqn\HgEnergyOne  $$
The desired change in energy
density with inclusion of the interactions is  $E(1) - E(0)$.
Since $\la \Psi(g) | \Psi(g)\ra = 1$,
$$\eqalign{
E(1)-E(0) & =\int_0^1 dg \,{d\over{dg}} \, E(g) \cr
&=\int_0^1 dg \, \la \Psi (g) |  {\d H(g)\over{\d g}} | \Psi (g)\ra \cr
&=\int_0^1 dg \, \la \Psi(g) | H^{(1)}+2g H^{(2)} | \Psi(g)\ra  ~~~. \cr}
   \eqn\HgEnergyTwo $$

The currents associated with  the fluctuation parts of the gauge fields are
defined by
$$\eqalign{
K^{\CS,\mu\nu} \, a_\nu &= K^{\mu\nu} \Big|_{\ep,\chi=0} \, a_\nu
   = \tilde j^\mu \cr
K^{\EM,\mu\nu} \, A_\nu &= K^{\mu\nu} \Big|_{\kappa=0} \, A_\nu
   = e \tilde j^\mu ~~.\cr}
  \eqn\gFieldEq  $$
$K^{\mu\nu}$ has been  defined in (\generalLagrangian).
In the theory descibed by the Hamiltonian $H(g)$,
$$\eqalign{
\tilde j^0&=g \, (\psi^\dagger \psi - \rhobar ) \cr
\tilde j^k&=-{ig\over 2m} \{ \psi^\dagger \Dbar_k \psi
 -(\Dbar_k \psi)^\dagger \psi \}
   -  {g^2\over m} (a^k + eA^k) \psi^\dagger \psi   ~~.  \cr}
          \eqn\gCurrentOne $$
Straightforward substitution of these currents yields
$$\int d\x \, \tilde j^\mu(x) \{ a_\mu(x)+eA_\mu(x) \} = gH^{(1)}+2 g^2 H^{(2)}
 ~~.
   \eqn\HgEnergyThree   $$
Thus
$$E(1)-E(0)=\int_0^1 {dg\over g} \int d\x \,
  \la \Psi(g) | \, \tilde j^\mu  (a_\mu +eA_\mu) \, | \Psi(g)\ra   ~~~.
     \eqn\HgEnergyFour  $$

If the gauge fields $A_\mu$ have (\generalGaugeField) as  a free
Lagrangian, their  equal-time commutation relations are given by (\commutator).
Elementary algebra shows that
$$\eqalign{
K^{\rho \mu}_x \la \T [A_\mu(x)A_\nu(y)] \ra
&= \la \T [ K^{\rho\mu} A_\mu(x)A_\nu(y)]\ra \cr
&\hskip .5cm  + g^{\rho \mu}
 \Big\{ 1 + \delta^{\rho 0} \big( {\ep\over \chi} -1 \Big) \Big\}
  \ep \delta(x_0-y_0) \, [\dot A_\mu(x) , A_\nu(y) ] \cr
&= \la \T [ K^{\rho \mu} A_\mu(x)A_\nu(y)]\ra
 + i {\delta^\rho}_\nu \delta^3 (x-y)  ~~.  \cr}
      \eqn\PropIdentityOne $$
The difference between the full and bare propagators, therefore, satisfies
$$K^{\rho \mu}_x    \Big\{ \la \T [A_\mu(x)A_\nu(y)]\ra
  - \la \T [A_\mu(x)A_\nu(y)]\ra_0 \Big\}
 = \la \T [ K^{\rho \mu} A_\mu(x)A_\nu(y)]\ra ~~~.  \eqn\PropIdentityTwo $$

Applying (\gFieldEq) and  (\PropIdentityTwo) to (\HgEnergyFour), one obtains
that
$$\eqalign{
&E(1) - E(0) \cr
&= \int_0^1{dg\over g}\int d\x \, \lim_{y\to x}
\bigg( K^{\EM,\nu \mu}_x \Big\{
\la \T [A_\mu (x)A_\nu (y)]\ra - \la \T [A_\mu (x)A_\nu (y)]\ra_0 \Big\} \cr
&\hskip 3cm + K^{\CS,\nu \mu }_x \Big\{
\la \T [a_\mu (x)a_\nu (y)] \ra -
\la \T [a_\mu (x)a_\nu (y)] \ra_0 \Big\} \bigg) \cr}
   \eqn\EnergyOne  $$
Fourier-transforming this expression yields, for the density,
$$\eqalign{
&\E(1)-\E(0) \cr
&= i\int_0^1 {dg\over g} \int {d^3p\over {(2\pi)^3}} \,
\bigg\{
\tr D_{0,EM}^{-1} \big( D[AA] - D_0^\EM \big)
 + \tr D_{0,CS}^{-1} \big( D[aa] - D_0^\CS \big) \bigg\} \cr
&= i\int_0^1 {dg\over g} \int {d^3p\over {(2\pi)^3}} \,
\Tr \hat D_0^{-1} \big( \hat D - \hat D_0 \big) \cr
&= i\int_0^1 {dg\over g} \int {d^3p\over {(2\pi)^3}} \,
\Tr \, \hat \Gamma \cdot {1\over \hat D_0^{-1} - \hat \Gamma}  ~~~. \cr}
           \eqn\EnergyTwo   $$
Here the two-by-two matrices $\hat D$, $\hat D_0$, and $\hat \Gamma$
are given by (\FullPropagatorOne) and (\FullPropagatorTwo).  The trace $\tr$
is taken over only Lorentz indices, whereas $\Tr$ is taken over both Lorentz
and
gauge field species.  We stress that the formula (\EnergyTwo) is exact,
no approximation being involved.

We have evaluated $\hat \Gamma_{\mu\nu}$ to the leading order in Section 5.
With the Hamiltonian (\TotalH), all $\Pi_k$'s there must be multiplied
by $g^2$.  In other words, $\hat \Gamma= {\rm O}(g^2)$.  The $g$-integral
in (\EnergyTwo), then, can be easily performed:
$$\eqalign{
\E(1)-\E(0)
&= -{i\over 2} \int {d^3p\over {(2\pi)^3}} \,
\Tr \ln \Big\{ 1 - \hat \Gamma|_{g=1} \hat D_0  \Big\} \cr
&= -{i\over 2} \int {d^3p\over {(2\pi)^3}} \,
\tr \ln \Big\{ 1 -  \Gamma^{(2)} (D_0^\CS + e^2 D_0^\EM) \Big\} ~~~. \cr}
     \eqn\EnergyThree  $$
$\Gamma^{(2)}$ is precisely the kernel evaluated in Section 5, which
represents the sum of one-loop diagrams.
The final expression in the above equation has  simple diagramatic
interpretation.  It represents the sum of a series of  ring diagrams connected
by
either Chern-Simons or electromagnetic fields.  (See Fig. 3.)
The `$\ln$' takes care of  combinatoric factors.

In the frame $p^\mu =(\omega, q,0)$
$$\Gamma^{\mu \nu}(p) =
\pmatrix {q^2\Pi_0 &\omega q\Pi_0 &iq\Pi_1 \cr
\omega q\Pi_0 &\omega^2\Pi_0 &i\omega \Pi_1 \cr
-iq\Pi_1 &-i\omega \Pi_1 &\Pibar_2 \cr}   ~~~.    \eqn\GammaMatrix $$
With a general propagator $D_0$ given by (\gaugePropagator),
$$-\Gamma D_0={1\over { \ep s-\kappa^2}}
\left( \matrix{
-\chi q^2 (\Pi_0+{\big \kappa\over \big s}\Pi_1)
   & \ep \omega q(\Pi_0+{\big \kappa\over \big s}\Pi_1)
   & iq (\ep \Pi_1+\kappa \Pi_0) \cr
-\chi \omega q(\Pi_0+{\big \kappa\over \big s}\Pi_1)
   & \ep\omega^2 (\Pi_0+{\big \kappa\over \big s} \Pi_1)
   & i\omega (\ep\Pi_1+ \kappa \Pi_0) \cr
i\chi q(\Pi_1+{\big \kappa \over \big s} \Pibar_2)
   & -i\ep \omega (\Pi_1+{\big \kappa \over \big s}\Pibar_2)
   & \ep \Pibar_2 +\kappa \Pi_1 \cr}    \right)  ~~,
   \eqn\GammaDone $$
where $s\equiv \ep \omega^2- \chi q^2.$
The electromagnetic part is obtained by taking the $\kappa \go 0$ limit in
the above formula.  For the Chern-Simons part we take the limit
$\epsilon, \chi \go 0$ with the ratio $\beta=\epsilon/\chi$ fixed.
We find
$$\eqalign{
-\Gamma &(e^2 D_0^\EM + D_0^\CS) = \left( \matrix{
  -q^2 a_1 & \omega q a_2 & iq c \cr
 -\omega q a_1 & \omega^2 a_2 & i\omega c \cr
 iq b_1 & - i\omega b_2 & d \cr} \right) \cr
&a_1 = {\chi\over \ep} {e^2\over s} \Pi_0 - {\beta\over \kappa u} \Pi_1   ~~,~~
a_2 = {e^2\over s} \Pi_0 - {1\over \kappa u} \Pi_1  \cr
&b_1 = {\chi\over \ep} {e^2\over s} \Pi_1 - {\beta\over \kappa u} \Pibar_2
{}~~,~~
b_2 = {e^2\over s} \Pi_1 - {1\over \kappa u} \Pibar_2  \cr
&c =  {e^2\over s} \Pi_1 - {1\over \kappa} \Pi_0  ~~,~~
d = {e^2\over s} \Pibar_2 - {1\over \kappa} \Pi_1  ~~, \cr}
    \eqn\GammaDtwo  $$
where $u=\omega^2 - \beta q^2$.    The $\beta$-dependence
in the expression for the Chern-Simons part should disappear in physical
quantities.

Indeed, for the energy density (\EnergyThree), we have
$$\eqalign{
\Delta \E &= \E(1)-\E(0) \cr
&= -{i\over 2} \int {d^3p\over {(2\pi)^3}} \,
\ln \det \Big\{ 1 -  \Gamma^{(2)} (D_0^\CS + e^2 D_0^\EM) \Big\} \cr
&= -{i\over 2} \int {d^3p\over {(2\pi)^3}} \,
\ln \bigg\{ 1 + {e^2\over s} \, \Pibar_2 + {e^2\over \ep} \, \Pi_0
 - {2\over \kappa} \, \Pi_1
+ \Big( {e^4\over \ep s} - {1\over \kappa^2 } \Big)
  \big( \Pi_0 \Pibar_2 - \Pi_1^2 \big) \bigg\}   ~.  \cr}
    \eqn\EnergyFour  $$
The shift in the energy density in pure Chern-Simons or pure
Maxwell theory is found as a special case:
$$\eqalign{
\Delta \E \big|_{\rm pure ~CS}
&= -{i\over 2} \int {d^3p\over {(2\pi)^3}} \,
\ln  {1\over \kappa^2} \,
 \Big\{ (\Pi_1 - \kappa)^2 - \Pi_0 \Pibar_2 \Big\}   \cr
\Delta \E \big|_{\rm pure ~EM}
&= -{i\over 2} \int {d^3p\over {(2\pi)^3}} \,
\ln  \bigg\{
\Big( 1 + {e^2\over \ep}\, \Pi_0 \Big) \Big( 1 + {e^2\over s}\, \Pibar_2 \Big)
  - {e^4\over \ep s} \, \Pi_1^2 \bigg\}  ~~.  \cr}
    \eqn\EnergyFive  $$

In terms of $\hat F_j(x,z) = F_j(x,iz)$ introduced in (\PiZero) and
(\PiOneTwo),
$$\eqalign{
\Delta \E
&= {1\over 4\pi^2 ml^4} \int_0^{\Lambda_x} dx \int_0^{\Lambda_z} dz \,
\ln \bigg\{ 1 + {e^2\over 2\pi ml^2 s} \, \hat F_2
 + {e^2 ml^2 \over 2\pi \ep} \, \hat F_0
 - {\ep(b_\tot) \over \pi \kappa} \, \hat F_1    \cr
&\hskip 5cm + {1\over (2\pi)^2}\Big( {e^4\over \ep s} - {1\over \kappa^2 }
\Big)
  \big( \hat F_0 \hat F_2 - \hat F_1^2 \big) \bigg\}   ~,  \cr}
    \eqn\EnergySix  $$
where $s= - (2\chi/l^2) x -(\ep/m^2l^4) z^2$.  Here we have introduced
ultraviolet cutoffs, $\Lambda_x$ and $\Lambda_z$,  supposing
that the model (\model) is an effective theory valid at low energies.
They are related to the momentum and frequency cutoffs by
$\Lambda_x= p_c^2 l^2/ 2$ and $\Lambda_z = ml^2 \omega_c$.

Expression (\EnergySix) contains many parameters.
Let us take  anyon superconductivity as  a typical example.
$\rhobar \sim 10^{14} \,{\rm cm}^{-2}$ so that
$b_\tot \sim$ 100$\,$T  and  $l \sim$ 10\AA.
The lattice spacing $a$ is about 5\AA.  For high $T_c$
superconductors one expects $p_c \sim a^{-1}$ and $\omega_c \sim 1\, {\rm eV}$.
The bare electron mass is $m = 2.6 \times 10^{10} {\rm cm}^{-1}$.
With these values $\Lambda_x \sim 2$, and $\Lambda_z \sim 2.6$.
Note that $\Lambda_z/\Lambda_x = 2 m\omega_c / p_c^2$ and that the effective
mass $m$ may be different from the bare electron mass.
 The coupling constant  $e^2/4\pi$ is $\sim \alpha/d$ where
$\alpha= 1/137$ is the fine structure constant and $d \sim 5{\rm \AA}$ is the
interplanar spacing.   This gives $e^2/4\pi \sim 1.5 \times 10^{5}
{\rm cm}^{-1}$, $e^2 l /4\pi  \sim 0.015$
and $e^2  ml^2/4\pi \sim 40$.  However,
in the Fradkin-Lopez picture of the fractional quantum Hall effect, there
are no cutoffs.  We shall come back to this point in Section 8.

In pure Chern-Simons theory
$$\eqalign{
\Delta \E
&= {\pi \rhobar^2 \over  m} \, R(\Lambda_x, \Lambda_z ;c)  \cr
R(\Lambda_x, \Lambda_z;c) &=
{1\over\pi} \int_0^{\Lambda_x} dx \int_0^{\Lambda_z} dz \,\ln \funR (x,z;c) \cr
\funR(x,z;c) &= ( 1  - c \hat F_1 )^2 -  c^2 \hat F_0 \hat F_2  ~~~,~~~
c={\ep(b_\tot) \over 2\pi \kappa} ~. \cr}
    \eqn\EnergyFormOne  $$
Here we have made use of the relation $\nu=1$ and $\rhobar=(2\pi l^2)^{-1}$.
$R(\Lambda_x, \Lambda_z;c)$ represents the RPA correction relative to the
mean field value.  It has  an important dependence on the number density
$\rhobar$  through $\Lambda_x$ and $\Lambda_z$.  A detailed numerical study is
presented in the following sections.

The behavior of $R(\Lambda_x, \Lambda_z;c)$ at large $\Lambda_x$ and
$\Lambda_z$
is analytically estimated with the aid of (\asymptoticF).
One finds that for $\Lambda_x, \Lambda_z > \Lambda_0 \ge 10$
$$\eqalign{
R(\Lambda_x, \Lambda_z;c) =& R(\Lambda_0, \Lambda_0;c)
+c\Big( {1\over 2}-{2\over \pi} \tan^{-1} {\Lambda_z\over \Lambda_x} \Big) \cr
& +c^2 \bigg\{ {1\over 2} \ln {\Lambda_z \over \Lambda_0} -
{1\over \pi} \int_1^{\Lambda_z/\Lambda_x} du ~
  {1\over u} \, \tan^{-1} u \bigg\}  ~. \cr}
    \eqn\EnergyLarge  $$
Notice that it grows logarithmically as $\Lambda$, or as the density $\rhobar$
gets smaller.

\vskip .5cm

\secno=7  \meqno=1

\leftline{\bf 7.  Anyon superconductivity}

A neutral anyon gas with pure Chern-Simons interactions has an application
to anyon superconductivity.\myref{\LaughlinOne-\HosotaniOne}  Previously, the
energy density in the  neutral anyon gas has been evaluated in the
Hartree-Fock approximation by Hanna, Laughlin and Fetter.\myref{\HLF,\Dai}
The correction to the energy density is $-{1\over 2}$ ($-{3\over 32}$) of
the mean field value in the $\nu=1$ ($\nu=2$) case,  independent of the
particle number density $\rhobar$.

It is known that  high $T_c$ cuprate superconductors become superconducting
only when a hole concentration $x_c$, namely the number of holes per
lattice plaquette, is in the range $0.05$ to $0.25$.   Within the context of
anyon superconductivity  it has been an unanswered question why
superconductivity is achieved only in a limited  range of $x_c$.

We shall show that in RPA the energy density is minimized exactly at a
number density in the range mentioned above.  At a lower or higher number
density the energy is increased, and the system is expected to lose
superconductivity.

Since $c= 1$ for  $\nu=1$ and $B_\ext=0$,
the total energy density of a neutral anyon gas is, from (\EnergyFormOne),
$$\eqalign{
\E(\rhobar)^\tot &= {\pi \rhobar^2 \over m} \,
    \big\{ 1 + R(\Lambda_x, \Lambda_z ;1)  \big\} \cr
R(\Lambda_x, \Lambda_z ;1) &=
{1\over \pi} \int_0^{\Lambda_x} dx \int_0^{\Lambda_z} dz \,
\ln \funR (x,z;1)    ~. \cr}
    \eqn\EnergyFormTwo  $$
The density dependence comes in through
$\Lambda_x=p_c^2/4\pi\rhobar$ and $\Lambda_z= m \omega_c/2\pi\rhobar$.
$R$ is evaluated numerically.

First we  depict, in  Fig.\ 4, the behaviour of the argument of the
logarithm, $\funR(x,z;1)$,  in the integrand.  It vanishes at the origin,
behaving as $2x + {1\over 12} x^2 + z^2$.  It is a smooth function, approaching
1 as $x$ or $z$ gets large.

As explained in the previous section, the momentum and frequency cutoffs
in anyon superconductivity are approximately $p_c \sim a^{-1}$ and
$\omega_c \sim 1 \, {\rm eV}$, where $a$ is the lattice spacing.  This gives
$\Lambda_z/\Lambda_x \sim 1$, but there remains an ambiguity.  We have
evaluated
$R$ as a function of $\Lambda_x$, or equivalently of
$(p_c a)^{-2} x_c =  \rhobar / p_c^2 = 1/(4\pi \Lambda_x)$,
with $\Lambda_z/\Lambda_x \equiv r$
fixed.

Fig.\ 5 (a) shows the behavior of $R$ for a wide range of $\Lambda_x$.
When $\Lambda_x, \Lambda_z > 10$ (small density) it grows logarithmically
($\sim {1\over 2} \ln \Lambda_x$) as expected from (\EnergyLarge).

In Fig.\ 5 (b), $R$ is plotted in the range $0 < x_c / (p_ca)^2 < 0.5$
where the minimum takes place.  We recognize that the minimum is located
around $x_c / (p_ca/\hbar)^2 = 0.05 \sim 0.06$ for a wide range of $r$ ($0.2 <
r
< 5$). If $p_c a/\hbar \sim 2$, $x_c^{\rm min} \sim 0.2$.  Although the value
of
$p_c$ is ambiguious,  it is safe to conclude that at a lower number density
the energy density sharply increases and therefore that the superconductivity
is lost at low densities.    The correction at the minimum ranges
from $-5$\% to $-25$\%, depending on $r$.
It is rather surprising, but also is encouraging,  that the
minimum occurs approximately at a number density where cuprate material is
 superconducting.

Certainly, more elaboration and detailed study of the anyon superconductivity
model is necessary in order to see if it can describe high $T_c$
superconductors
or yet-to-be-discovered new material.  We leave it for future investigation.

\vskip .5cm

\secno=8 \meqno=1

\leftline{\bf 8.  Fractional quantum Hall effect}

In the Jain-Fradkin-Lopez picture of the fractional quantum Hall
effect, electrons in an external magnetic field with a filling factor
$\nu_\ext$
are described by composite fermions interacting with, in addition to
the external magnetic field, Chern-Simons gauge fields, as was explained in
Section 2.\myreff{\Jain-\Simon}   Our previous result is valid for the total
filling factor $\nu = n= 1$, which applies, from (\FQHEvalue), to a sequence
$\nu_\ext = 1/(2p \pm 1)$.

In the electron picture the lowest Landau level has an energy
$|eB_\ext|/2m$ so that the mean field energy density is
$$\E_{\rm MF}^{\rm electron} = {1\over \nu_\ext} \cdot
  {\pi\rhobar^2\over m} ~~~.  \eqn\electronMFenergy   $$
In the composite fermion picture the lowest Landau level has an energy
$|b_\tot|/2m$  so that the mean field energy for $\nu=1$ is
$$\E_{\rm MF}^{\rm composite} =
  {\pi\rhobar^2\over m} ~~~.  \eqn\compositeMFenergy   $$
Clearly $\E_{\rm MF}^{\rm composite} < \E_{\rm MF}^{\rm electron}$.
There appears a large discrepancy, by a factor of $\nu_\ext^{-1}$.

Since the transformation from the original electron system to the
Chern-Simons-fermion system is exact, the discrepancy observed above
is merely an artifact of the mean field approximation.  We are going to show
that quantum fluctuations give a big correction.  It is expected that
if one includes all corrections, the energy densities in the two systems should
be exactly the same.

Without loss of generality we take $eB_\ext >0$ as before.  From (\FQHEvalue)
one finds that for $\nu=1$
$$\eqalign{
\nu_\ext &= {1\over 2p \pm 1} \quad {\rm where~} \pm = \ep(b_\tot) \cr
c &= \mp 2p  ~. \cr}
         \eqn\FQHEkappa  $$

As shown by Halperin, Lee and Read \myref{\Halperin} and by Simon and
Halperin \myref{\Simon},  RPA gives a substantial correction
to the excitation energy spectrum.   In the absence of the Coulomb interaction,
the spectrum is determined by zeros of $\det D_\CS^{-1} \propto
(\Pi_1 - \kappa)^2 - \Pi_0 \Pibar_2$.  In terms of $F_j$
$$(1 - c F_1)^2 - c^2 F_0 F_2 =0~~.   \eqn\FQHEspectrum $$
At a zero momentum $x = {1\over 2} q^2 l^2 =0$,
$$\eqalign{
F_0(0,y) &= F_1(0,y) = 1 + F_2(0,y)  \cr
&= {1\over 1 - y^2 - i\ep} \cr}  \eqn\FQHEspectrumTwo  $$
Substitution of (\FQHEspectrumTwo) into (\FQHEspectrum) yields
$$y = ml^2 \omega = |c-1| = 2p  \pm 1 = {1\over \nu_\ext} ~,
   \eqn\FQHEspectrumThree  $$
or
$$\omega(q=0)^{\rm RPA} =  {1\over \nu_\ext} {2\pi\rhobar\over m} ~~.
  \eqn\FQHEspectrumFour  $$
This is exactly the Landau gap in the original electron picture.
The RPA correction yields a factor $\nu_\ext^{-1}$ compared
with the mean field value in the Chern-Simons picture.

The energy density in RPA is given by
$$\eqalign{
\E(\rhobar)^{\rm RPA} &= {\pi \rhobar^2 \over m} \,  ( 1 + R ) \cr
R(\Lambda_x, \Lambda_z; c) &=
{1\over \pi} \int_0^{\Lambda_x} dx \int_0^{\Lambda_z} dz \,
\ln \funR (x,z;c)   ~. \cr}
    \eqn\EnergyFormThree  $$
We have evaluated $R$ as a function of $\Lambda_x$ for various values of
$r=\Lambda_z/\Lambda_x$ and $c$.    There correspond two values of $c$
to each $\nu_\ext$.  For instance, $\nu_\ext= {1\over 3}$ (${1\over 5}$)
corresponds to $c=-2$ and $+4$ ($-4$ and $+6$).  The function $\funR (x,z;c)$
in
(\EnergyFormThree) is depicted in Fig.\ 6.

For a large cutoff $\Lambda_x$, $R \sim {1\over 2} c^2 \ln \Lambda_x$ as
follows from (\EnergyLarge).  Its global behaviour is depicted in Fig.\ 7.
For a given $\nu_\ext$, the RPA correction exceeds the saturation value
$\nu_\ext^{-1} -1$ at around $\Lambda_x = 10 \sim 100$.  This implies that
RPA is not accurate for large momenta and frequencies.

In the light of the equivalence between the original electron picture and
the composite fermion  picture the cutoffs must be removed.  In one respect
our result shows that higher order corrections beyond RPA become crucial
in the computation of the energy density.   It  asserts that quantum
fluctuations give a large correction of order $\nu_\ext^{-1}$.

\vskip .5cm

\secno=9 \meqno=1

\leftline{\bf 9. Summary}

We have examined the nonrelativistic Maxwell-Chern-Simons gauge theory relevant
in the quantum Hall effect and anyon superconductivity.
We have derived compact integral representations for the polarization tensors
$\Gamma^{\mu\nu}(p)$ or $\Pi_k(p)$.    Quantum fluctuations
give important contributions to the energy density of
the system, which  is most conveniently expressed  in terms of the gauge field
propagators.   The energy density was evaluated in RPA.

One of the important consequences in RPA is that there results a
non-trivial dependence of the energy density on the particle number density.
In particular, in anyon superconductivity, the ratio of the energy
density in RPA to that in the mean field approximation is minimized at a hole
concentration  $x_c = (0.05 \sim 0.06) (p_c a/\hbar)^2$.
In light of the application
to high $T_c$ superconductors this is extremely intriguing and encouraging.
Typical high $T_c$ cuprates become superconducting for
$0.05 < x_c < 0.25$.  Neither the mean field approximation nor
the Hartree-Fock approximation have been successful in explaining why
superconductivity is achieved only in a limited range of $x_c$.

In the application to the fractional quantum Hall effect, the RPA correction to
the energy density in the Fradkin-Lopez picture diverges as the cutoffs
are removed.  The divergence comes from large momenta and frequencies where
RPA is expected to break down.  We may conclude that higher order quantum
fluctuations give substantial corrections to the energy density, presumably
restoring the equivalence between the original electron picture and the
composite fermion picture.

In this paper we have concentrated on the case of a unit filling $\nu=1$ with
respect to the total magnetic field $b_\tot=\rhobar/\kappa + eB_\ext$.  The
generalization to arbitrary integer filling $\nu=n$ is important both for
the anyon superconductivity and for the fractional quantum Hall effect.  It
is left for future investigation.

\vskip .5cm

\leftline{\bf Acknowledgements}

This work was supported in part
by the U.S.\ Department of Energy under contract no. DE-AC02-83ER-40105,
and by the Japan Society for the Promotion of Science.
One of the authors (Y.H.) is grateful to the
National Laboratory for High Energy Physics (KEK) in Japan for its hospitality
where a part  of this work was done.
D.W.\ would
like to thank K.A.\ K\"atzchen-Tondra and M.\ Groblewski-Higgins
for encouragement and support.

\vskip .5cm

\leftline{\bf References}
{\ninerm
 \immediate\closeout\reffile
	\input refs.tmp   
   
}

\vskip .5cm

\baselineskip=14pt plus 1pt minus 1pt
\parindent=32pt
\parskip=8pt

\leftline{\bf Figure captions}

\item{Fig. 1}  One-loop corrections $\Gamma^{\mu\nu}$ defined in
(\FullPropagatorTwo) and (\PiFormulaTwo).  Three kinds of vertices are given
by the last three terms in (\LagrangianFive).

\item{Fig. 2}  The behaviour of $\hat F_j(x,z)$ in (\FFormula).
(a) $\hat F_1(x,z)$,
(b) $\hat F_2(x,z)$ and
(c) $1+\hat F_2(x,z)$.
All are smooth functions of $x$ and $z=iy$. They start out with unity at the
origin and approach zero asymtotically for large $x$ or $z$ (\asymptoticF).

\item{Fig. 3}  The energy density in RPA.  See Eq.\ (\EnergyThree).
A doubly dashed line represents the sum of Chern-Simons and electromagnetic
field propagators.

\item{Fig. 4}   Plot of the argument of the natural logarithm in
(\EnergyFormTwo) as a function of $x$ and $z=iy$. It vanishes at the origin and
approaches unity asymtotically  for large $x$ or $z$. (See text for more
details on the asymtotic  behaviour.) This is related to the RPA energy density
of a neutral anyon gas.  Note that conventional order of $x$ and $z$ has been
switched for visual  clarity.

\item{Fig. 5}   Magnitude of the RPA correction to the mean field energy
density
$R(\Lambda_x,\Lambda_z;1)$
for a neutral anyon gas normalized to the mean field value for various choices
of the lattice cutoff $\Lambda_x$ and $\Lambda_x$ (\EnergyFormTwo). The solid
line  denotes cutoff ratio $r=\Lambda_z/\Lambda_x={1\over 5}$, the dashed
line denotes  $r=1$ and the dotted line denotes $r=5$.
(a) Plotted as a function of $\Lambda_x$ for fixed values of $r$ on a
logarithmic scale. Linearity for large values of $\Lambda_x$ is due to the
logarithmic growth of $R$ with cutoff (\EnergyLarge). Departure from linearity
at very
large values of $\Lambda_x$ is understood to be a numerical artifact. Note
that the minima for $r=1$ and $r=5$ are not resolved. They are resolved in the
following figure. Numerical estimates of $R$ are accurate to about 5\%.
(b) Alternative presentation of Fig. 5(a) around the minima. $R$ has been
plotted as a function of $x_c(p_ca/\hbar)^{-2} = 1/(4\pi\Lambda_x)$
(scaled hole concentration),
where $p_c$ and $a$ are the momentum cutoff and
lattice spacing, respectively. The
minima  are adequately resolved and their locations are more or less
independent
of  the cutoff ratio $r$. For large values of concentration, $R$ approaches
zero.  We note intriguing similarity with the cuprate superconductors.
Numerical
estimates of $R$ are accurate to about 10\%.

\item{Fig. 6}  Plot of the argument of the natural logarithm in
(\EnergyFormThree) for  $\nu_{\ext}={1\over 3}$ as functions of $x$ and $z=iy$.
Corresponding values of the  constant $c$ are $-2$ and $+4$. Both start
with $\funR=9$ at the origin.
See text for  more details on the asymtotic behaviour.
Note that conventional order of  $x$ and $z$ has been switched for visual
clarity.  (a) Plot for $c=+4$.
(b) Plot for $c=-2$.

\item{Fig. 7}   Magnitude of the RPA correction to the mean field energy
density
$R(\Lambda_x,\Lambda_z;c)$ for a quantum Hall system normalized to the mean
field value for various  choices of the lattice cutoff $\Lambda_x$ and
$\Lambda_z$ (\EnergyFormThree). $R$ is  plotted as a function of $\Lambda_x$
for
fixed values of  $r=\Lambda_z/\Lambda_x$ on a logarithmic scale. Linearity for
large values of  $\Lambda_x$ is due to the logarithmic growth of $R$ with
cutoff
(\EnergyLarge).  $R$ is always positive in all cases.
Departure from linearity at very large values of $\Lambda_x$ is
understood to be  a numerical artifact. Numerical estimates of $R$ are accurate
to about 5\%. (a) Behaviour for a $c=4$ ($\nu_{\ext}={1\over 3}$) system. The
solid line denotes  cutoff ratio $r={1\over 5}$, the dashed line denotes
$r=1$ and the dotted line  denotes $r=5$.
(b) Behaviour for various choices of external filling fraction with $r=1$.
The solid line denotes $c=-2$ ($\nu_{\ext}={1\over 3}$), the dashed line
denotes  $c=+6$ ($\nu_{\ext}={1\over 5}$), the dotted line denotes $c=-4$
($\nu_{\ext}={1\over 5}$) and the dash-dotted line denotes $c=+4$
($\nu_{\ext}={1\over 3}$).

\vfil\eject

\end